# Design of a 64-bit SQRT-CSLA with Reduced Area and High-Speed Applications in Low Power VLSI Circuits

CH. Pallavi [1], C. Padma[2], R. Kiran Kumar[3], T. Suguna[4], C. Nalini[5]

[1]Associate Professor, Department of ECE, Sri Venkateswara College of Engineering (SVCE), Tirupati, A.P, India.
[2]Associate Professor, Department of ECE, Sri Venkateswara College of Engineering (SVCE), Tirupati, A.P, India.
[3]Associate Professor, Dept. of. ECE, MITS, Tirupati, A.P, India.
[4]Assistant Professor, Dept. of. ECE, Panimalar Engineering College, Chennai, India.
[5]Assistant Professor, Dept. of. ECE, Mohan Babu University (MBU), Tirupati, A.P, India.

pallavi.ch@svcolleges.edu.in
ORCID ID: 0000-0002-3283-8460

**ABSTRACT**

The main areas of research in VLSI system design include area, high speed, and power-efficient data route logic systems. The amount of time needed to send a carry through the adder limits the pace at which addition can occur in digital adders. One of the quickest adders, the Carry Select Adder (CSLA), is utilized by various data processing processors to carry out quick arithmetic operations. It is evident from the CSLA's structure that there is room to cut back on both the area and the delay. This work employs a straightforward and effective gate-level adjustment (in a regular structure) that significantly lowers the CSLA's area and delay. In light of this adjustment Square-Root Carry Select Adder (SQRT CSLA) designs with bit lengths of 8, 16, 32, and 64. When compared to the standard SQRT CSLA, the suggested design significantly reduces both area and latency. Xilinx ISE tool is used for Simulation and synthesis. The performance of the recommended designs in terms of delay is estimated in this study using the standard designs. The study of the findings indicates that the suggested CSLA structure outperforms the standard SQRT CSLA.

**Keywords:** CSLA (Carry Select Adder), SQRT CSLA (Square Root Carry Select Adder), VLSI, Xilinx.

## 1 INTRODUCTION

Power-efficient and fast data route logic system design is among the most important fields of research in VLSI. With regard to digital adders, the propagation time of a carry through the adder limits the adder's speed. Any digital system, whether it be for digital signal processing or control, needs to be able to add. A digital system's ability to operate quickly and precisely is largely dependent on how well its resident adders execute. Because adders are widely used in other fundamental digital operations like subtraction, multiplication, and division, they are also a crucial part of digital systems. Therefore, enhancing the digital adder's performance would significantly improve the way binary operations are carried out inside a circuit made up of these blocks.

Analysing a digital circuit block's power dissipation, layout area, and operation speed allows one to determine how well it performs. The two primary areas of research on VLSI system design are reduced area and fast data route logic systems. High-performance processors and systems have always required addition and multiplication to operate at high speeds. The amount of duration required for a carry through the adder to propagate limits the pace at which addition can occur in digital adders. In a basic adder, the sum for each bit position is produced in a sequential manner only after the preceding bit position and a carry has propagated into the subsequent position. There are several computational systems that employ the CSLA.

In an elementary adder, each bit sum is formed by serially after adding the preceding bit, the carry created by this addition is transferred to the following bit.. The Carry Select Adder (CSLA) generates numerous carry bits and selects a carry for the required output in a number of computer systems to alleviate the problem of carry propagation delay. However, because the CSLA uses multiple pairs of Ripple Carry Adders (RCA) to produce a partial sum and carry by taking carry data into account, and multiplexers select the final total and carry (mux), the CSLA is not area-efficient. High-performance processors and systems have always required addition and multiplication to operate at high speeds.

Design to separately produce many carriers, one carry is selected to generate the aggregate in order to mitigate the issue of carry propagation latency. To get the final amount, it employs Cin=0 and Cin=1. However, because it employs numerous pairs of RCA to generate limited sum or carry by taking carry input into consideration, the regular CSLA is not area or speed efficient. The multiplexers determine the final sum and carry (mux). The area will grow as a result of using two separate RCAs, which will increase the delay.

The fundamental idea of the suggested Utilizing n-bit BEC is the task to increase addition speed in order to solve the aforementioned issue. To further increase speed and decrease the delay, the logic can be replaced by means of Cin is 1 in RCA. To speed up the addition process by reducing area and delay. the standard CSLA can be replaced with BEC. The primary benefit of the BEC logic is that it uses fewer logic gates than the Full Adder (FA) structure since fewer gates are needed.

Used CSLA is to deduce the complexity of delay in carry creation and then selecting the sum [1]. The CSLA, on the other hand, greater area, and it employs multiple blocks of RCA to calculate the carry and sum by first treating the carry inputs as 0 and 1, after which the multiplexer selects the amount of sum or carry [2–6]. The fundamental concept of this paper is that AND/OR gates generates carry and is carried to the subsequent step, whereas only sum is chosen by the multiplexer. Figure 1 shows a block diagram of regular CSLA.

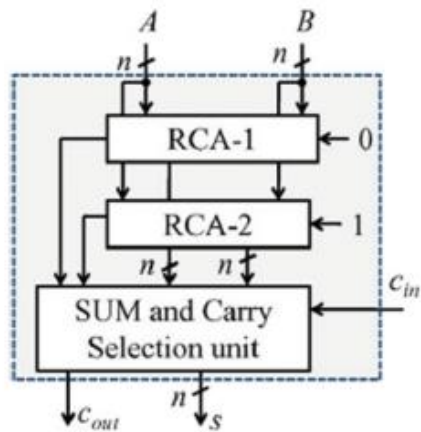

**Fig. 1** Architecture of a R-CSLA

The carry input is first taken as 0 to generate the sum and carry, and then it is again taken as 1 to generate the sum and carry. The multiplexer will choose the sum and carry appropriately since it receives the actual input carry-in the pick line at the end. As a result, ripple carry adder blocks are utilized for computation in each step. In the diagram, A and B stand for the input values, while S and Cout stand for the output sum and carry.

The structure of this document is as follows. The previous carry select adder works are presented in Section 2. In Section 3, the suggested CSLA is described and the area and power reduction are assessed. Section 4 analyses the proposed CSLA's execution details and results, and wraps up the work in Section 5.

## 2 RELATED WORKS

The amount of time it takes to propagate carry signal through digital adders limits the adder's speed. To address the issue of carry propagation delay, the R-CSLA was developed. It selects the appropriate carry output and sum based on the preceding carry's value, after independently producing several carries. The following are the current carry select adder works:

**M. Vinod Kumar Naik et al [7],** proposed a CSA designed for rapid speed and low-powered VLSI applications. This work suggests a novel logic formulation for CSLA and offers a way to remove the all-logic operations seen in standard CSLA. This scheme differs from the traditional CSLA. Xilinx ISE was used to synthesize the suggested CSLA, and Xilinx Power Analyzer was used to analyse the power.

**Nilkantha Rooj et al [8]**, provided an analysis of the internal design of the traditional CSLA and other carry select adder designs in addition to developing a new design. In this work, carry selection operates ahead of the sum computation for each stage, but it does so in a different way than a traditional CSLA. The carry selects the unit's logic has been minimized by using a certain bit pattern found in the carry out results produced by the distinct carry values (Cin is 0 & Cin is 1) in the suggested CSLA. A novel and efficient design for CSLA is suggested, based on this enhanced logic formulation.

**Gagandeep Singh and Chakshu Goel [9]**, explained EX-OR gates are the fundamental building blocks and 8-bit adder is designed using a 3-T EX-OR gate. In comparison to standard CSLA and modified CSLA, the suggested CSLA has fewer transistors, consumes less power, and has a lower power-delay product (PDP). This research uses a straightforward method to improve the EX-OR gate's efficiency. The 3-T EX-OR gate used in the proposed design results in a significant reduction in the no. of MOSFETs. This suggested 8-bit CSLA has a power decrease of 27.7% and 21.7%, respectively.

**R. Sakthivel and G. Ragunath [10]**, Three blocks, HSCG, FCG, and FSG, were proposed for a low power, high speed, efficient CSLA. In comparison to the current CSLA, the HSCG is faster and requires less space and power thanks to its low-complexity Boolean expression design. In comparison to the current design, a 64-bit suggested CSLA delivers an average 43% power savings and an average 25% reduction in area consumption.

**S. Allwin Devaraj et al [11]**, Proposed a Pass Transistor Logic (PTL) technology is being used in the design of CSLA to further reduce power and space. PTL reduces the no. of transistors needed for each logical circuit, which lowers the carry select adder's size and power consumption. Tanner EDA tool is used to run the simulation. Moreover, superfluous transistors and propagation delay can be decreased. In order to obtain less area and power, the PTL approach is employed to minimize the complexity at the transistor level.

**Dr. D. B. Kadam et al [12]**, provided a straightforward method for using the BEC-1 architecture to minimise the SQRT-CSLA delay and area. The decrease of gates and LUT's is made possible by the BEC in the structure. This modified approach is a good substitute for adder implementation in many data processors as it reduces both area and delay. The goal of this work is to minimise the Area, delay and power of the CSLA architecture.

**S. Balaprasad et al [13]**, developed a new logic formulation for CSLA and proposed a method that removes all of the unnecessary logic operations included in the traditional CSLA.

Unlike the traditional method, the suggested technique schedules the final-sum calculation before the CS operation.

**B. Ramkumar et al [14]**, presented with only a minor increase in delay, the suggested design outperforms the standard SQRT-CSLA. The study of the findings indicates that the suggested CSLA structure outperforms the standard SQRT CSLA.

**Nagulapati Giri et al [15]**, Through comparison of metrics like as area, delay, and power consumption, the effectiveness of all design techniques has been examined. High-speed multiplication, arithmetic logic units, sophisticated microprocessor architecture, and other applications can benefit from the high efficacy design. The gpdk180 library was used, and Cadence Virtuoso Analog Design Environment was used to simulate each architecture.

**Bagya Sree Auvla et al [16]**, High-performance, low-power, and area-efficient Multi-standard wireless receivers, biomedical instruments, and portable and mobile devices are among the growing number of applications for VLSI systems. By employing AOI logic to simplify the BEC and RCA units, the gate counts are decreased. The suggested CSLA is put into practice for various word sizes. Low power applications like digital signal processing, multipliers, filter ERROR accuracy, and design may benefit from the suggested architecture.

**Kadaru Prasanna et al [17]**, The CSLA is frequently used to address this. A major breakthrough in system design of digital adders in VLSI is presented by the suggested improvements of including a BEC and a Kogge-Stone adder into the CSLA architecture.

**Nelanti Harish et al [18]**, In contrast to the traditional method, Prior to the final-sum calculation, a new logic formulation for the CSLA operation is scheduled. Carry words that correspond to input-carry '0' and '1' produced by the CSLA using the suggested technique adhere to a particular bit pattern that is utilized for the CS unit's logic optimization. An ideal design for the CS and CG units is produced as a result. These optimized logic units are used to create an effective CSLA design.

**S. Muminthaj et al [19]**, The suggested design uses less power than the standard adder circuits. This proposed study offers a straightforward method to lower the CSLA architecture's area, power consumption, and latency. The traditional Carry Select Adder's drawbacks include higher power consumption, increased chip area use, and a significant latency. The outcomes are compared in terms of area, delay, and power consumption. When compared to the previous model, the D-Flip-flop-based Improved CSLA turns out to be the High Speed and Low Power CSLA.

## 3 METHODOLOGY AND IMPLEMENTATION

The difference between this architecture and a standard 64-bit SQRT CSLA is that a BEC is used to substitute the RCA with Cin=1 out of the two available RCAs in a group. One of the features of this BEC is that it can carry out operations that are compared with BEC logic. Fig. 3.1 illustrates the 64-bit SQRT CSLA's modified diagram. A bit more bits are needed for BEC logic than for RCA logic. Along with being divided into several groups of bits of varying sizes, the updated block diagram also has appropriate muxes, BECs, and ripple carry adders for each group. Group 0 comprises a single RCA that receives a lower significant bit as input, carries it. Finally, the selected input arrives after the RCA and BEC for the remaining groups.

Consequently, the output from mux, sum1, and the results calculated by BEC and RCA, depend on mux. sum2 is dependent upon mux and c1. The onset time of the selected mux input for the remaining components is consistently higher than the arrival time of the BEC data inputs.

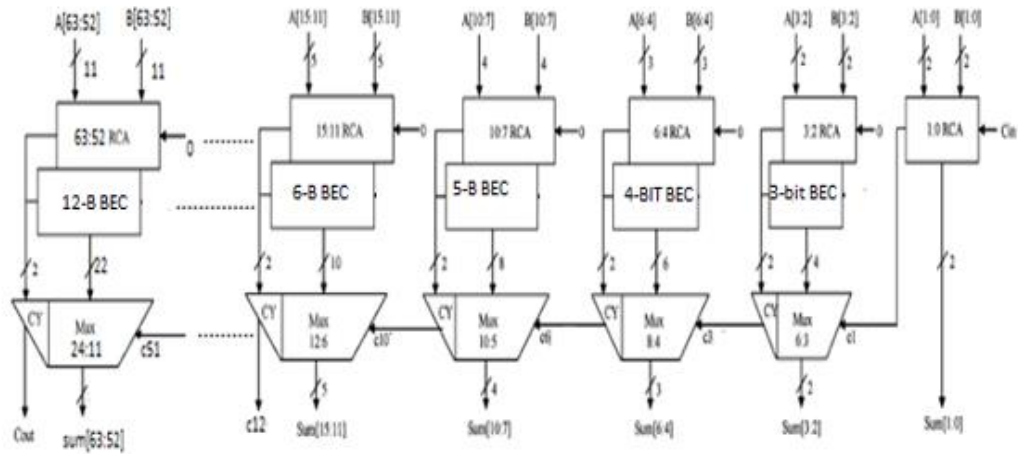

**Fig. 3.1**: Modified 64-bit SQRT CSLA block diagram

The fundamental operation of 6-bit addition, which consists of 12-bit mux, 6-bit data, and BEC logic (6-bit), is depicted in Fig. 3.2. The addition process is carried out for Cin = 0 and
Cin = 1. Ripple carry adders are used for addition when Cin=0, and 6-bit BECs are used for operations when Cin=1 (replacing the RCA). Based on the previous group's carry in signal, the resultant is chosen. The previous group's Cin signal and mux delay determine the overall delay. Figure 3.2, depict the construction of a 6-bit BEC with 12:6 MUX.

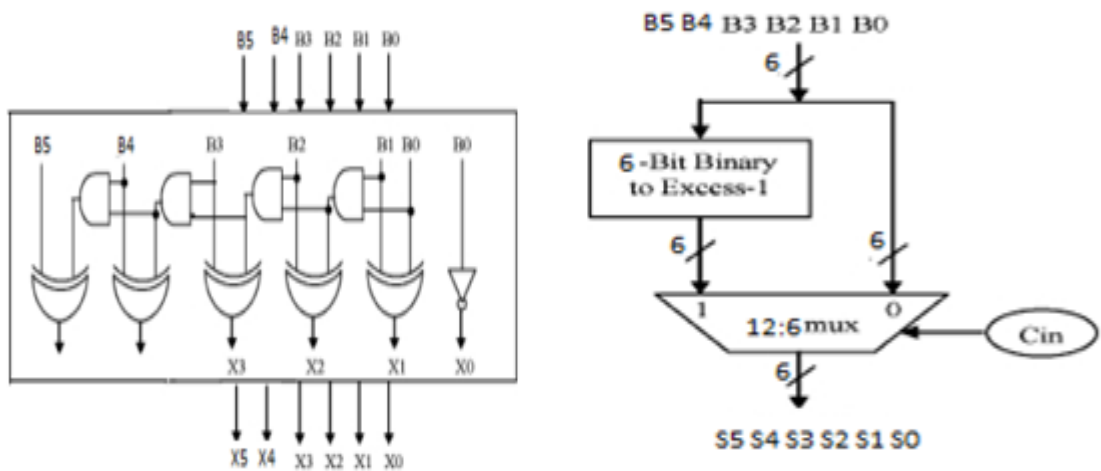

**Fig. 3.2:** Structure of a 6-bit BEC

### 3.1 ESTIMATION OF A 64-BIT MODIFIED SQRT CSLA

Figure 3 illustrates the architecture of the suggested 64-bit SQRT CSLA that uses BEC for Ripple Carry Adders (RCA) with CIN = 1 to maximize power and area.1. There are five groups within the framework. Fig. 3.3 displays the area estimation and delay for group 5. These are the steps that lead to the evaluation:

1) For CIN = 0, the group 2 [see Fig. 3.3(a)] has one 2-b RCA with 1 FA and 1 HA. A 3-b BEC is utilized in place of an additional 2-b RCA with CIN = 1, adding one to the output from the 2-b
2) RCA. The arrival time of selection input c1 [time (t) = 7] of the 6:3 mux is later than the s2 [t = 4] and earlier than the s3 [t = 9] and c3 [t = 10], according to Table I's delay values. As a result, mux and s3 and partial c3 (input to mux) and mux, respectively, are required for the sum3 and final c3 (output from mux). The sum2 is dependent upon mux and c1.
3) The mux selection input arrival time for the remaining groups is consistently higher than the arrival time of the BEC data inputs. As a result, the arrival time of the mux selection input and the mux delay determine the remaining groups' delays.
4) The following formula is used to get group 2's area count:
   (FA + HA + Mux + BEC) = 43 gates
   XOR = 10(2 * 5) (BEC) Mux = 12(3 * 4) FA = 13(1 * 13) HA = 6(1 * 6) AND = 1, NOT = 1.
5) Similarly, Table 3.2 lists and evaluates the anticipated maximum latency and area of the other groups of the updated SQRT CSLA.

Table 3.2: Modified SQRT CSLA groups' area count and delay

| Delay (in ns) | Area (in gates) | Group |
|---|---|---|
| 13 | 43 | 2 |
| 16 | 61 | 3 |
| 19 | 84 | 4 |
| 22 | 107 | 5 |

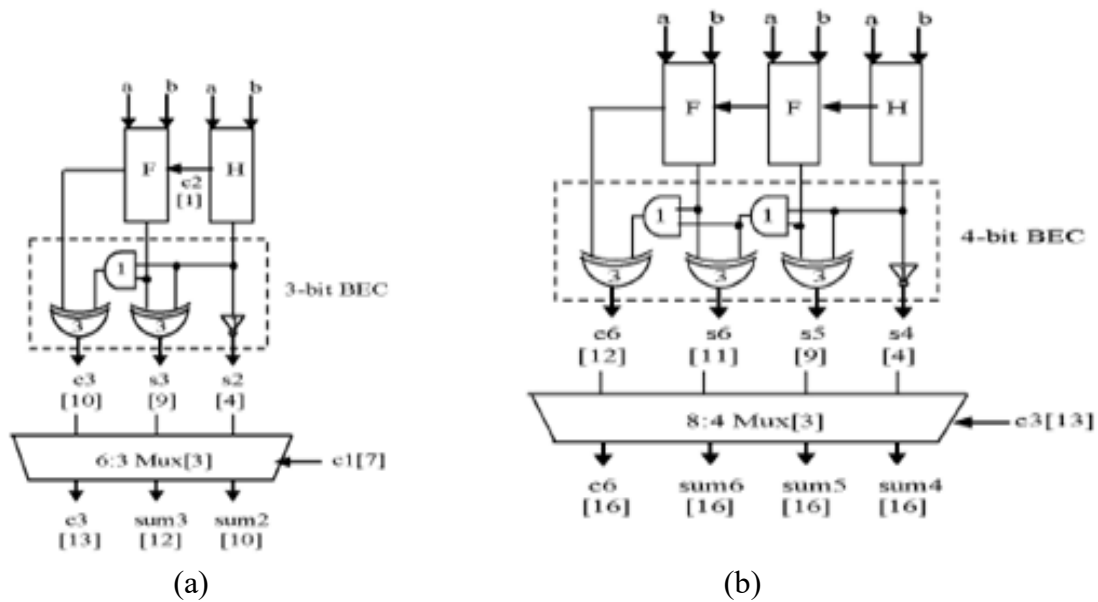

(a)            (b)

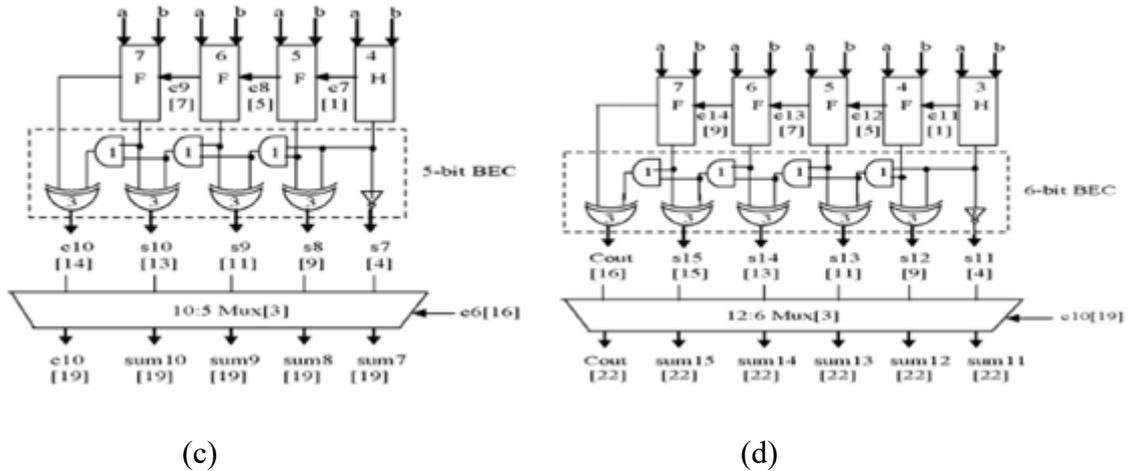

(c)                  (d)

**Fig 3.3** Figure 3.3: Delay and area analysis of the suggested SQRT CSLA Groups (a) and (b), groups (c) and (d) and (e)

## 4 RESULTS AND DISCUSSION

This work mainly concentrates on the design of "Gate level modification" of a 64-bit SQRT CSLA for reduced area application. The Model Sim and Xilinx ISE tools are used for simulation and synthesis process. The simulation and synthesis results are shown below. The simulation result for the 64-bit SQRT CSLA for reduced area application in Model Sim (shown in Fig.4.1) is as follows,

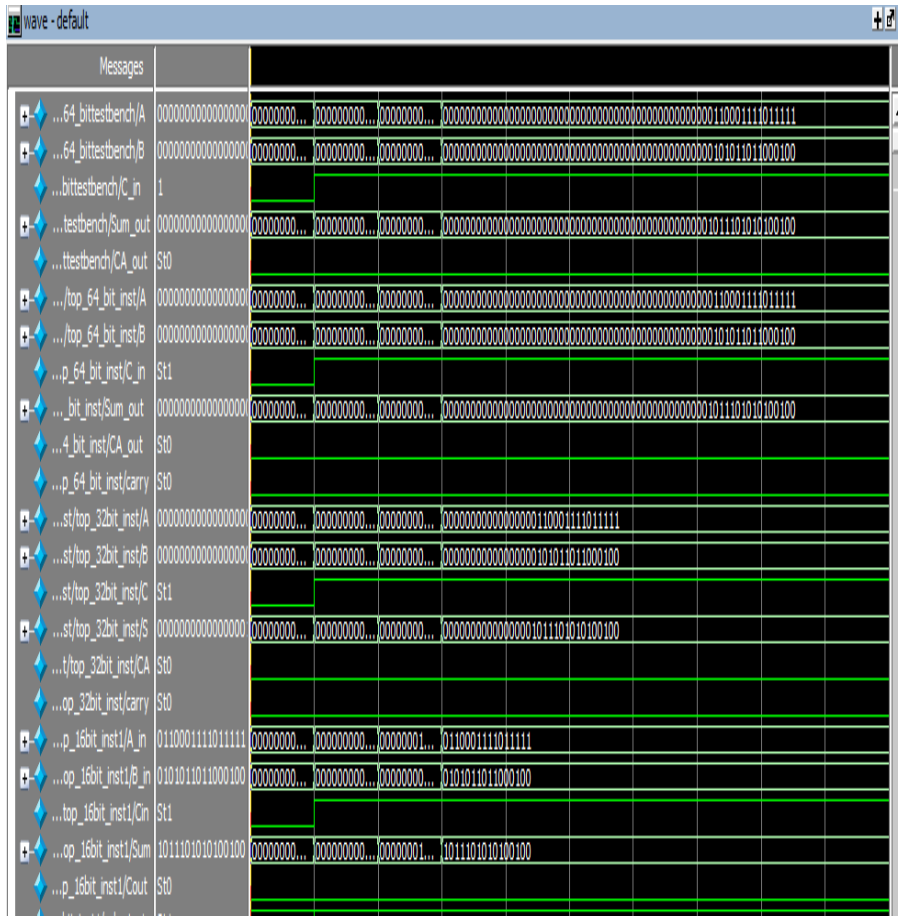

**Fig. 4.1:** Simulation result for the 64-bit SQRT CSLA

Input Data:
  A=64'd25567; B=64'd22212; Cin=1'b1;
Output Data:
  Sum=64'd47780; Cout=1'b0;

The synthesis of 64-bit SQRT CSLA for reduced area applications in Xilinx yields Top level schematic, RTL schematic and Technology View (shown in Fig. 4.2,4.3 and 4.4) as follows,

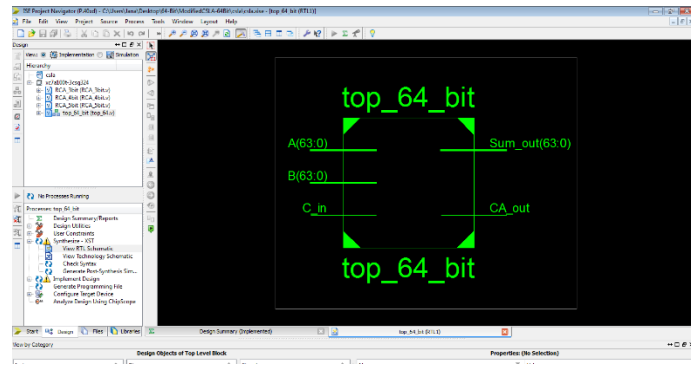

**Fig.4.2:** Top level schematic

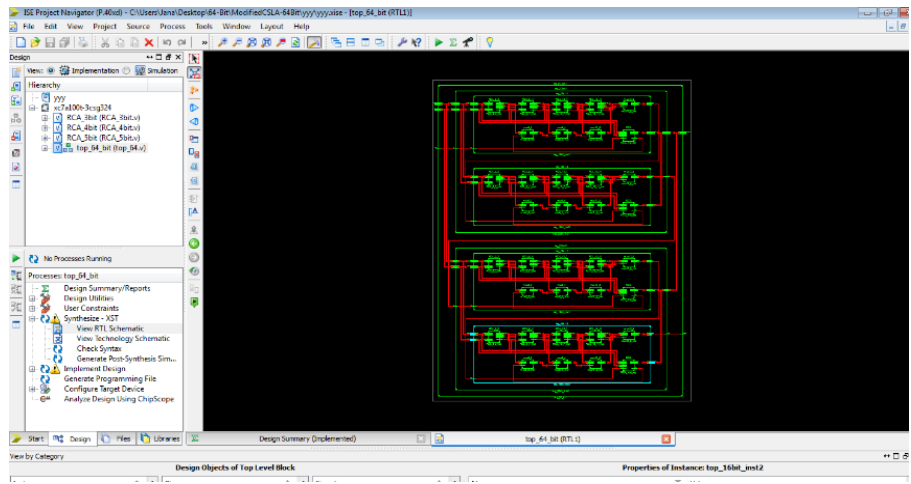

**Fig.4.3:** RTL schematic

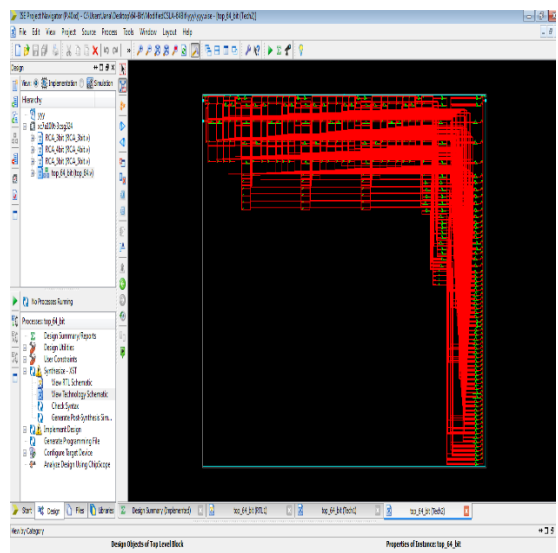

**Fig.4.4:** Technology schematic

The comparison table for regular & modified 64-bit SQRT CSLA is shown below.

**Table 4.1: Comparison Table for Regular & Modified 64-bit SQRT CSLA**

| Parameter | Regular | Modified |
|---|---|---|
| No. of bits | 64-bit | 64-bit |
| Type of Algorithm | RCA | BEC |
| No. of LUT's | 162 | 135 |
| No. of Adders | 240 | 96 |
| Area (in gates) | 1352 | 1169 |
| Delay (in ns) | 20.461ns | 17.596ns |
| Speed | Low | High |

It is evident from the foregoing that, in comparison to the ordinary approach, the 64-bit modified method has a lower delay. Thus, the new method greatly reduces both the area and the latency.

## 5 CONCLUSION AND FUTURE WORK

This study presents an effective method for decreasing the size and delay of 64-bit SQRT CSLA architecture. All that needs are used to get the number of gates to be reduced in the structure and is to swap out the RCA for BEC. This work's fewer gates provide a substantial help in terms of decreased area, latency, and overall power. When compared to the standard 64-bit SQRT CSLA architecture, the modified architecture has a smaller area and less delay, as demonstrated by the comparison findings. The findings therefore indicate that the area and time would decrease when the improved method is used. In hardware implementation of VLSI, the 64-bit SQRT CSLA modified architecture is utilized since it is high speed, low area and delay, straightforward, and effective. This technology is utilized to execute several algorithms, such as FFT, FIR, and IIR, in a variety of applications, including multipliers and DSP.

The idea can be further developed for a larger bit count in the future. 512, 128, 256, and so forth.

**REFERENCES**

[1] T.Y. Ceiang, M.J. Hsiao, Carry-select adder using single ripple carry adder. Electron. Lett. 34(22), 2101–2103 (1998)

[2] B. Ramkumar, H.M. Kittur, Low-power and area-efficient carry select adder. IEEE Trans. Very Large Scale Integr. VLSI Syst. 20(2), 371–375 (2012)

[3] S. Manju, V. Sornagopal, An efficient SQRT architecture of carry select adder design by common boolean logic, in Proceedings of VLSI ICVENT, pp. 1–5 (2013)

[4] K.B. Sindhuri, Implementation of regular linear carry select adder with binary to excess-1 converter. Int. J. Eng. Res. 4(7), 346–350 (2015)

[5] K.M. Priyadarshini, N.V. Kiran, N. Tejasri, T.C. Anish, Design of area and speed efficient square root carry select adder using fast adders. Int. J. Sci. Technol. Res. 3(6), 133–138 (2014)

[6] B.K. Mohanty, S.K. Patel, Area–delay–power efficient carry-select adder. IEEE Trans. Circuits Syst. II 61(6), 418–422 (2014

[7] M. Vinod Kumar Naik and Mohammed Aneesh. Y, "Design of Carry Select Adder for Low-Power and High-Speed VLSI Applications", 978-1-4799-6085-9/15/$31.00 ©2015 IEEE.

[8] Nilkantha Rooj, Snehanjali Majumder and Vinay Kumar, " A Novel Design of Carry Select Adder (CSLA) for Low Power, Low Area, and High-Speed VLSI Applications", Springer Nature Singapore Pte Ltd. 2018 J. K. Mandal et al. (eds.), Methodologies and Application Issues of Contemporary Computing Framework, https://doi.org/10.1007/978-981-13-2345-4_2.

[9] Gagandeep Singh and Chakshu Goel, "Design of Low Power and Efficient Carry Select Adder Using 3-T XOR Gate", Hindawi Publishing Corporation Advances in Electronics Volume 2014, Article ID 564613, 6 pages http://dx.doi.org/10.1155/2014/564613.

[10] R. Sakthivel and G. Ragunath, "Low power area optimized and high-speed carry select adder using optimized half sum and carry generation unit for FIR filter", Journal of Ambient Intelligence and Humanized Computing, May 2021, Springer 2020. Doi: 10.1007/s12652-020-02062-3

[11] S. Allwin Devaraj, R. Helen Vedanayagi Anita and S. Abirami, "Design of Carry Select Adder with Reduced Area and Power", International Journal of Advanced Information Science and Technology (IJAIST) ISSN: 2319:2682 Vol.4, No.2, February 2015 DOI:10.15693/ijaist/2015.v4i2.46-50.

[12] Dr. D. B. Kadam, Dr K K Pandyaji, and Dr. Kazi Kutubuddin Sayyad Liyakat, "Implementation of Carry Select Adder (CSLA) for Area, Delay and Power Minimization", TELEMATIQUE Volume 21 Issue 1, 2022 ISSN: 1856-4194, pp: 5461 – 5474.

[13] S. Balaprasad, and M. Jeyalakshmi, "Area Efficient Carry Select Adder with Low Power", SSRG International Journal of Electrical and Electronics Engineering (SSRG-IJEEE) – volume 2 Issue 2 Feb 2015.

[14] B. Ramkumar and Harish M Kittur, "Low-Power and Area-Efficient Carry Select Adder", IEEE Transactions on Very Large-Scale Integration (VLSI) Systems, VOL. 20, NO. 2, FEBRUARY 2012, DOI: 10.1109/TVLSI.2010.2101621.

[15] Nagulapati Giri, and Muralidharan D, "A Survey on Various VLSI Architectures of Carry Select Adder", International Journal of Innovative Technology and Exploring Engineering (IJITEE) ISSN: 2278-3075 (Online), Volume-8 Issue-5, March 2019.

[16] Bagya Sree Auvla, R. Kalyan, "Low Power, Area Efficient & High-Performance Carry Select Adder on FPGA", International Journal of Innovative Research in Computer and Communication Engineering, Vol. 3, Issue 5, May 2015, DOI: 10.15680/ijircce.2015.030507.

[17] Kadaru Prasanna, Dewkathe Divya, and Godise Badrinath, "Design of High Speed and Low Power Carry Select Adder", International Journal of Research Publication and Reviews, Vol 4, no 6, pp 1472-1478 June 2023.

[18] Nelanti Harish, V. Sarada and T. Vigneswaran, "High Speed, Low Power and Area Efficient Carry-Select Adder", I J C T A, 9(15), 2016, pp. 7167-7174.

[19] S. Muminthaj, S. Kayalvizhi, and K. Sangeetha, "Low Power and Area Efficient Carry Select Adder Using D-Flip Flop", International Journal of Science and Research (IJSR), Volume 8 Issue 11, November 2019, 10.21275/ART20202769.

# Authors:

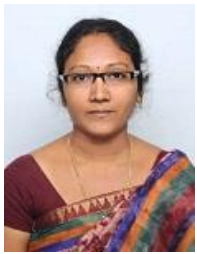

Dr. CH Pallavi https://orcid.org/0000-0002-3283-8460
Scopus ID: 58695425600
Received her Ph.D. from Sri Venkateswara University, Tirupati, M. Tech from JNTUA, Anantapur in VLSI System Design as specialization and B. Tech from SKIT, Srikalahasti, JNTUA, Anantapur and currently working as Associate Professor in the Dept. of ECE, Sri Venkateswara College of Engineering (SVCE), Tirupati. She is having more than 13 years of teaching experience and has more technical publications in National /International journals and conferences. She has filed and published National and International Patents and few of them granted too. She is acting as a reviewer and program committee member for many International and National conferences and journals. Her research areas and publications include signal & Image processing, Communications, VLSI design, and Underwater Wireless Communication System. She has **NPTEL** certifications on IOT, Cloud Computing, Computer Networks and Internet Protocol etc.

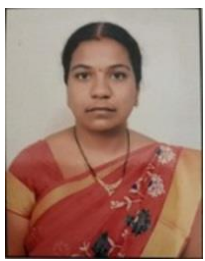

Dr. C Padma  https://orcid.org/0000-0002-8880-1624
Scopus ID: 57215831376
Web of Science Researcher ID: IUN-3464-2023
A Associate Professor in the Department of ECE,SVCE Tirupati. She has completed her M.Tech in VLSI System Design in 2013 and Ph. D in VLSI and Signal Processing in 2023 from Jawaharlal Nehru Technological University, Ananthapuramu. She has published papers in national and international journals. Her area of Interest includes VLSI and Signal processing, IOT and Embedded Systems.

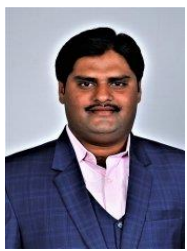

Dr. R Kiran Kumar https://orcid.org/0000-0003-0671-9641
Scopus ID: 57212111551
Web of Science Researcher ID: AAP-2132-2021
A Assistant Professor in the Department of ECE,MITS Madanapalle. He has completed B.Tech from SVCET-Jawaharlal Technological University Ananthapur and M.Tech from SJCET-Jawaharlal Technological University Ananthapur and Ph.D from Vel Tech Rangarajan Dr.Sagunthala R&D Institute of Science and Technology, Avadi, Chennai, India.

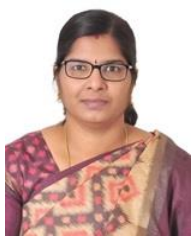

Dr. T. Suguna is a distinguished academic and researcher with a Ph.D. in Low Power VLSI Design, M.Tech in VLSI SD and B.Tech in Electronics and Communication Engineering. She has teaching experience, of 10 years and currently working as Assistant Professor G1 at Panimalar Engineering College, Chennai. Her commitment to continuous learning is evident through her successful completion of multiple NPTEL certification courses, including topics like CMOS Digital VLSI Design, Embedded System Design with ARM, and System Design through Verilog, achieving high scores and earning Elite medals.

She is also a prolific researcher, with numerous publications in Scopus-indexed journals, covering advanced topics in VLSI design, energy-efficient circuits, adiabatic logic, and digital signal processing. In addition to her research, she has actively participated in seminars, workshops, and faculty development programs, contributing to her broad expertise in 5G design, IoT applications, and high-performance VLSI architectures

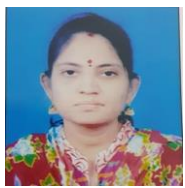

Nalini Chekuri received her M.Tech degree in VLSI System Design from JNTUA Ananthapuramu in 2010, and is now currently working as Assistant Professor in ECE department at Sri Venkateswara College of Engineering, Tirupati. Her area of interests includes Low Power VLSI Architectures, Signal Processing and IOT. She can be contacted at email: nalini.chekuri02@gmail.com